\documentclass[sigconf]{acmart}
\usepackage{graphicx,caption,subcaption,booktabs,tabularx,array,colortbl,longtable,makecell,multirow}
\usepackage{float,placeins,microtype,enumitem}
\newcolumntype{Y}{>{\raggedright\arraybackslash}X}
\newcolumntype{Q}[1]{>{\raggedright\arraybackslash}p{#1}}
\AtBeginEnvironment{table}{\hyphenpenalty=10000\exhyphenpenalty=10000}
\AtBeginEnvironment{table*}{\hyphenpenalty=10000\exhyphenpenalty=10000}
\newcommand{\duo}{\textsc{DuoSketch}}
\newcommand{\dquote}[1]{\textit{``#1''}}
\definecolor{U1}{HTML}{AC5139}
\definecolor{U2}{HTML}{806126}
\definecolor{U3}{HTML}{38785C}
\definecolor{U4}{HTML}{3B6F9E}

\definecolor{C1}{HTML}{8F1838}
\definecolor{C2}{HTML}{D52E3F}
\definecolor{C3}{HTML}{EF7666}
\definecolor{N4}{HTML}{244A80}
\definecolor{N6}{HTML}{2878BD}
\definecolor{N5}{HTML}{4260CD}
\definecolor{N1}{HTML}{537F9E}
\definecolor{N2}{HTML}{79B8E5}
\definecolor{N3}{HTML}{648DE4}
\definecolor{S1}{HTML}{C49A16}
\definecolor{S2}{HTML}{DBAE24}
\definecolor{S3}{HTML}{EBC64C}
\DeclareRobustCommand{\scode}[1]{\textcolor{S#1!75!black}{\textbf{S#1}}}
\definecolor{Phase1}{HTML}{276749}
\definecolor{Phase2}{HTML}{338A58}
\definecolor{Phase3}{HTML}{4AA66A}
\definecolor{Phase4}{HTML}{76B88B}
\DeclareRobustCommand{\singlephasebadge}[1]{\textcolor{Phase#1!65!black}{\textbf{Phase #1}}}
\DeclareRobustCommand{\phasebadge}[1]{\ifstrequal{#1}{2--3}{\singlephasebadge{2}\,\singlephasebadge{3}}{\singlephasebadge{#1}}}

\DeclareRobustCommand{\ccode}[1]{\textcolor{#1!85!black}{\textbf{#1}}}

\DeclareRobustCommand{\nshort}[1]{\textcolor{N#1!75!black}{\textbf{N#1}}}

\setcopyright{none}
\acmDOI{}
\acmISBN{}
\begin{document}
\raggedbottom
\title{\duo{}: How Pairs Navigate Challenges in AI-Supported Collaborative Design Ideation}

\author{Weiyan Shi}
\email{weiyanshi6@gmail.com}
\orcid{0009-0001-6035-9678}
\affiliation{
  \institution{Singapore University of Technology and Design}
  \city{Singapore}
  \country{Singapore}
}

\author{Darryl Lim}
\email{owlycodes@gmail.com}
\orcid{0009-0007-8173-2199}
\affiliation{
  \institution{Singapore University of Technology and Design}
  \city{Singapore}
  \country{Singapore}
}

\author{Geraldine Quek}
\email{geraldine\_quek@sutd.edu.sg}
\orcid{0000-0003-2864-3860}
\affiliation{
  \institution{Singapore University of Technology and Design}
  \city{Singapore}
  \country{Singapore}
}

\author{Kenny Tsu Wei Choo}
\email{kennytwchoo@gmail.com}
\orcid{0000-0003-3845-9143}
\affiliation{
  \institution{Singapore University of Technology and Design}
  \city{Singapore}
  \country{Singapore}
}

\begin{abstract}
Generative AI offers new resources for collaborative design ideation, yet how designers work with it as a shared concept develops remains underexplored. We developed \duo{}, integrating a shared canvas and live transcripts with a separate AI for each member of a pair. An exploratory qualitative study with 12 pairs identified three challenges: unresolved fit between AI proposals and shared design purposes, shared meanings not expressed in AI responses, and agreed changes missing from later outputs. We found that pairs navigated these challenges through bridging work in two directions: \emph{AI $\rightarrow$ Shared Design} (reworking features, developing uses, and assessing fit with shared requirements) and \emph{Shared Design $\rightarrow$ AI} (supplying materials, explaining meanings, and communicating decisions). We discuss how future AI systems can support the joint exploration and articulation of design ideas.
\end{abstract}

\begin{teaserfigure}
\centering
\includegraphics[width=\textwidth]{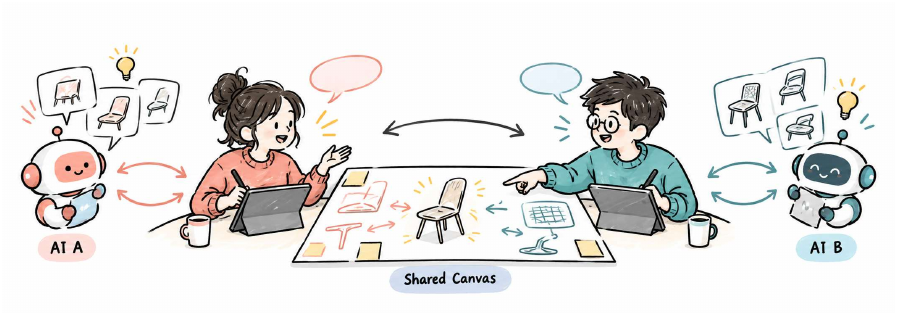}
\caption{Collaborative design with \duo{}. Pairs sketch and discuss on a shared canvas, with a separate AI available to each collaborator. They can submit canvas materials and live discussion transcripts to AI and bring generated images back to the canvas.}
\Description{Illustration of two collaborators discussing chair concepts around a shared canvas, each using a tablet and interacting with an AI. Coral and teal distinguish the two sides. Curved arrows connect each collaborator with an AI, and a two-way arrow connects the collaborators. The robot characters symbolise the AIs.}
\label{fig:setting}
\end{teaserfigure}

\maketitle
\section{Introduction}

Design ideation involves exploring problems and possible solutions, with design intentions taking shape through that exploration~\cite{schon1983reflective,dorst2001coevolution}. In collaborative design, this process unfolds through discussion, sketching, and interaction with shared artefacts~\cite{tang1991observational}. Generative AI introduces another source of design material into these activities. Understanding its role in collaboration requires examining how designers work with it over the course of developing a shared concept.

The collaborative setting matters because a design is developed through contributions from people who may begin with different ideas and interpretations. Sketches and conversation help collaborators explore possibilities and establish enough shared understanding to proceed~\cite{clark1991grounding,tang1991observational,vanderlugt2002functions}. What they want to make can change as they encounter and discuss alternatives. AI-generated material enters this open-ended process: its relevance extends beyond the exchange in which it was produced to the design activity the collaborators undertake together. This motivates studying AI use in relation to the development of a shared concept over time.

Multimodal design tools help designers express and explore concepts through sketches, language, and generated representations~\cite{kazi2017dreamsketch,subramonyam2024envisioning,shi2026drawing}. Shared AI workspaces support group ideation and collective interpretation~\cite{shaer2024brainwriting,he2024canvas,zhang2025ladica}, while visual collaboration tools support communication through AI-generated references~\cite{ko2022wetoon,chung2023artinter}. These developments provide ways to incorporate AI into collaborative activity. They also motivate a process-oriented question: what happens as collaborators continue working with AI beyond an initial proposal, combining ideas, exploring a changed situation, and preparing a concept they can present together?

This question is especially relevant when collaborators have separate AI interactions alongside a shared workspace. Personal and group activity already coexist in collaborative work~\cite{scott2004territoriality}; providing each collaborator with AI introduces further opportunities for individual exploration within a joint task. Related studies have begun examining shared and separate AI access in collaborative programming and software design~\cite{daryanto2026triadic,jackson2026collaborative}. In visual design ideation, collaborators also work through speech, sketches, and generated images. A setting that combines these shared resources with separate AIs offers an opportunity to examine how individual AI use participates in an evolving collaborative process.

Less is understood about the difficulties pairs encounter as they develop a concept through this combination of shared work and individual AI exchanges, and how they respond to those difficulties. Following a concept across successive activities can reveal demands that an isolated generation task would leave unexamined. An empirical account of these encounters and responses can inform how future AI systems support collaborative design ideation throughout concept development.

To investigate this process, we developed \duo{}, which integrates a shared canvas and live discussion transcript with a separate AI for each member of a pair. Pairs can submit shared materials to AI and bring generated images back to the canvas. We conducted an exploratory qualitative study with 12 pairs as they synthesised, adapted, and consolidated design concepts. Two questions guide the study:
\begin{itemize}[leftmargin=*]
\item \textbf{RQ1:} What challenges do pairs encounter during AI-supported collaborative design ideation?
\item \textbf{RQ2:} How do pairs navigate these challenges as their shared design develops?
\end{itemize}

We make three contributions:
\begin{itemize}[leftmargin=*]
\item An empirical account of three challenges pairs encounter in connecting shared design with AI: unresolved fit between proposals and shared purposes, shared meanings missing from responses, and agreed changes absent from later outputs.
\item An account of bridging work through six navigation strategies in two directions: developing AI proposals within shared design, and communicating shared design to AI. These strategies show how navigation involves both shaping shared intentions and expressing them.
\item Design implications for supporting joint judgement, joint expression, and communication of evolving design context, grounded in observed interactions and pairs' suggestions for support.
\end{itemize}

\section{Related Work}
\label{sec:related}

\subsection{Generative Tools for Exploring and Developing Design Concepts}

Design-support tools help people give form to concepts that are still taking shape. Earlier systems such as DreamSketch combined freehand sketching with generative design algorithms: a designer sketched a design context, explored alternative three-dimensional solutions, and inspected them within that context~\cite{kazi2017dreamsketch}. DesignScape offered both refinements of a current layout and alternatives for exploring a different arrangement~\cite{odonovan2015designscape}. May AI? adapted inspirational suggestions through interaction with the designer~\cite{koch2019mayai}. These systems connect computational assistance to ongoing visual reasoning as a concept takes shape.

Recent multimodal generative tools extend this approach through conversational and image-based interaction. Baudoux et al. combine text-based and sketch-based conversations to support conceptual design, allowing designers to express and develop concepts through both language and drawing~\cite{multimodal_conceptual_design2025}. SketchConcept supports sketch-based concept composition for product design using multimodal large language models~\cite{sketchconcept2026}. In car exterior design, Lee et al. connect three-dimensional sketching with two-dimensional generative AI, enabling designers to develop visualisations from their sketches while working across viewpoints~\cite{lee2025car}. These systems make the designer's artefacts part of the input to generation. They provide ways to communicate form and composition that would be difficult to express through a textual prompt alone.

Speech provides another way to express intentions alongside a sketch. TalkSketch combines freehand drawing with real-time speech input to support design ideation~\cite{shi2026talksketch}. Shi et al. subsequently examine spontaneous speech accompanying sketches, comparing sketch-only input with sketches augmented by concurrent speech transcripts for intent alignment~\cite{shi2026drawing}. These studies motivate attention to meaning expressed across visual and verbal materials. Related work also examines changing human--AI roles during sketch-based ideation~\cite{shi2026taxonomy}. With pairs, these intentions develop through shared discussion and sketching alongside AI interaction.

Other tools make generation itself more open to revision. PromptPaint uses paint-like interactions to steer image generation, Prompt\-Charm combines prompting with multimodal refinement, and Promptify helps users explore alternative prompts and their outputs~\cite{chung2023promptpaint,wang2024promptcharm,brade2023promptify}. Beyond visual design, AI Chains exposes intermediate steps for inspection and editing~\cite{wu2022aichains}; IdeaSynth supports evolving and composing research-concept facets, while CoQuest examines how interaction with an AI supports research-question development~\cite{pu2025ideasynth,liu2024coquest}. Across these tasks, generation provides material for an unfolding process.

The resulting interaction also changes what designers want to make. Designers must anticipate what an AI can produce and work out how to express their intentions, a difficulty described as the gulf of envisioning~\cite{subramonyam2024envisioning}. Multimodal input expands their options, but does not make this work straightforward. Wadinambiarachchi et al.'s comparison of text, sketch, and combined prompting found that participants' modality preferences did not simply follow the benefits of sketching for exploratory work~\cite{reflection_in_action2026}. Their earlier study also found that AI-generated visual inspiration could increase fixation and reduce the diversity of subsequent concepts~\cite{wadinambiarachchi2024fixation}. Generative tools therefore support both expressing an intention and encountering material that may redirect it.

In collaborative design, a generated proposal becomes material for joint interpretation. An image that helps one designer express a concept may become the basis on which both collaborators choose what to pursue. Our study follows how pairs interpret AI contributions together and express the meanings and decisions developed through that shared work.

\subsection{AI Support for Shared Design Work}

Group ideation systems bring AI assistance into activities that people already undertake together. AI-Augmented Brainwriting incorporates LLM support into group concept generation and evaluation~\cite{shaer2024brainwriting}. He et al.'s Collaborative Canvas explores LLM-supported group ideation in a virtual workspace, including participants' concerns about ownership and preserving human contributions~\cite{he2024canvas}. LADICA uses a large shared display to help co-located teams organise and analyse their concepts, drawing on spoken discussion to support collective work~\cite{zhang2025ladica}. In these systems, the shared activity provides a focal point for encountering and discussing AI contributions.

Visual collaboration also has a substantial design-support lineage. ImageSense combines individual and shared moodboard workspaces with several forms of intelligent assistance~\cite{koch2020imagesense}. SemanticCollage connects visual material with semantic labels to support searching and articulating design concepts~\cite{koch2020semanticcollage}. We-toon helps webtoon writers communicate revisions to artists through generated and locally manipulated reference images, addressing the difficulty of expressing visual feedback in words~\cite{ko2022wetoon}. Artinter supports artist--client communication through a joint moodboard that connects verbal concepts with visual examples, using AI-assisted search and generation to develop these materials~\cite{chung2023artinter}. These tools show how AI-supported artefacts can help people communicate across differences in expertise and interpretation. 

Personal and shared activity coexist in collaborative workspaces. Studies of tabletop collaboration distinguish personal, group, and storage territories~\cite{scott2004territoriality}, while work on co-creativity foregrounds communication and feedback between people and AI~\cite{rezwana2025communication}. Daryanto et al. compare collaborative programming with a shared AI and with an AI for each participant, alongside individual human--AI programming~\cite{daryanto2026triadic}. In a workshop study of collaborative software design, Jackson et al. identify context drift between separate LLM instances~\cite{jackson2026collaborative}. These studies draw attention to the relationship between AI access and shared activity. In multimodal design ideation, this relationship also involves shared sketches, spoken explanations, and generated references that pairs interpret and reuse.

Recent preprints extend this question to other settings. Zha et al. compare individual and shared GenAI mentors in classroom group work~\cite{zha2026configurations}, while Cho et al. examine AI-assisted interpretation of collaborative discussion in groups with shared history~\cite{cho2026sharedhistory}. These studies address AI access and interpretation in classroom work and group discussion. We examine these issues during multimodal design ideation, where pairs develop concepts through conversation, sketches, and generated representations.

In \duo{}, pairs discuss and sketch on a shared canvas while consulting separate AIs. We examine how they connect these activities as they combine, adapt, and consolidate a design concept.

\subsection{Shared Representations and Joint Design Judgement}

Designs develop through interaction with representations. Accounts of reflection-in-action and the co-evolution of design problems and solutions show how designers discover what matters through making and interpreting design material~\cite{schon1983reflective,dorst2001coevolution,goldschmidt1991dialectics}. Human--AI alignment offers a related perspective on how AI contributions respond to and shape people's intentions: Shen et al. consider both adaptation of AI to humans and adaptation of humans in interaction with AI~\cite{shen2025alignment}. In design, AI proposals can prompt revisions to both a representation and the intention guiding it.

With two collaborators, developing a shared design also requires establishing enough shared understanding to proceed. Grounding describes how people accomplish this through interaction~\cite{clark1991grounding}, while workspace awareness explains the importance of seeing what another person is doing~\cite{gutwin2002awareness}. Studies of collaborative drawing show that speech and sketching work together in communicating and developing design concepts~\cite{tang1991observational,vanderlugt2002functions}. Experiments on shared visual information further distinguish support for grounding from support for awareness of the task situation~\cite{gergle2013visual}. Access to the same image is therefore not equivalent to agreement about its significance. Carlile's distinction between transferring, translating, and transforming knowledge offers a complementary perspective on why moving material across a boundary may also require interpreting or revising it~\cite{carlile2004boundaries}. Shared intentions may be expressed across discussion, sketches, and generated artefacts. Distributed cognition provides a complementary account of how such work is accomplished across people, representations, and tools~\cite{hollan2000distributed,zhang1994representations}.

The alternatives available for comparison also matter. Dow et al. found benefits of parallel prototyping for design exploration, and subsequently showed how sharing multiple prototypes affected pairs' exploration, critique, and outcomes~\cite{dow2010parallel,dow2011sharing}. With generative AI, suggestions can influence more than the artefact: opinionated writing assistants affected users' expressed views, and language-model-assisted writing could reduce content diversity~\cite{jakesch2023opinionated,padmakumar2024diversity}. These findings concern different tasks and mechanisms, but motivate attention to what remains visible when collaborators consider AI contributions. A meta-analysis likewise shows that the value of combining human and AI work varies across tasks and comparison conditions~\cite{vaccaro2024combinations}. 

Grounding, joint interpretation, and human--AI adaptation provide foundations for understanding how pairs develop designs with AI. Our analysis follows how shared intentions are shaped through joint responses to AI proposals and then expressed in AI requests. It connects changes in the design itself---such as developing a feature's use---with the work of communicating materials, meanings, and decisions. This relationship between design development and expression guides our account of bridging work.

\section{Method}
\label{sec:method}

We conducted an exploratory qualitative study in Singapore with 12 pairs (24 participants) undertaking a collaborative design ideation task with \duo{}. Each member of a pair used a separate AI chatbot while both worked on the same canvas and could see the live discussion transcript. We followed how pairs developed AI proposals together and communicated shared materials, meanings, and decisions to AI. We use \emph{shared design} to mean the concepts, requirements, and intended uses that pairs articulated and took up through explicit agreement, joint modification, or use following discussion. An \emph{AI history} comprises the prompts, attachments, and outputs accumulated through one chatbot during the session.

\subsection{Participants}

We recruited 24 participants (13 female, 11 male; ages 18--44, with 20 aged 18--24) through university mailing lists and local channels. All participants were fluent in English, and all design discussions, prompts, and reflection interviews were conducted in English. Every participant had completed at least one design-related course, representing backgrounds across design, engineering, and the humanities. Nine pairs were previously acquainted; three were formed for the study. None had prior experience with \duo{}. Table~\ref{tab:participants} reports participant background, self-rated sketch confidence (7-point scale), weekly generative AI use, and pair familiarity.

The study received institutional ethics approval (institution redacted for review). Participants gave informed consent and each received US\$7.80 upon completing the study. We identify pairs as P1--P12; P1a and P1b, for example, denote the two collaborators in Pair 1. Each AI is identified by its collaborator, for example P1a's AI and P1b's AI.

\begin{table*}[htbp]
\caption{Participant characteristics. Sketch confidence is self-rated on a 7-point scale; AI use is the reported weekly frequency of using generative AI for design. F denotes female and M male.}
\label{tab:participants}
\centering\small
\begin{tabular}{@{}llllccc@{}}
\toprule
\rowcolor{black!9}
\textbf{Participant} & \textbf{Age} & \textbf{Gender} & \textbf{Background} & \shortstack{\textbf{Sketch}\\\textbf{confidence}} & \shortstack{\textbf{AI use}\\\textbf{(/week)}} & \shortstack{\textbf{Previously}\\\textbf{acquainted}} \\
\midrule
\rowcolor{black!4}
\textbf{P1a} & 21--24 & F & Recent graduate (non-design) & 5 & 5--10 & Yes \\
\rowcolor{black!4}
\textbf{P1b} & 35--44 & F & Professional (design) & 7 & 5--10 & Yes \\
\addlinespace[2pt]
\textbf{P2a} & 21--24 & M & Design student & 4 & 5--10 & Yes \\
\textbf{P2b} & 21--24 & F & Non-design student & 4 & 5--10 & Yes \\
\addlinespace[2pt]
\rowcolor{black!4}
\textbf{P3a} & 21--24 & M & Non-design student & 4 & 5--10 & Yes \\
\rowcolor{black!4}
\textbf{P3b} & 18--20 & M & Non-design student & 5 & 5--10 & Yes \\
\addlinespace[2pt]
\textbf{P4a} & 25--30 & M & Professional (design) & 4 & 5--10 & No \\
\textbf{P4b} & 35--44 & M & Professional (non-design) & 5 & $>$10 & No \\
\addlinespace[2pt]
\rowcolor{black!4}
\textbf{P5a} & 18--20 & F & Design student & 5 & 5--10 & No \\
\rowcolor{black!4}
\textbf{P5b} & 21--24 & F & Non-design student & 4 & 5--10 & No \\
\addlinespace[2pt]
\textbf{P6a} & 25--30 & F & Professional (design) & 5 & 5--10 & No \\
\textbf{P6b} & 21--24 & F & Former art student & 6 & 5--10 & No \\
\addlinespace[2pt]
\rowcolor{black!4}
\textbf{P7a} & 21--24 & M & Design student & 6 & 5--10 & Yes \\
\rowcolor{black!4}
\textbf{P7b} & 21--24 & F & Design student & 5 & 5--10 & Yes \\
\addlinespace[2pt]
\textbf{P8a} & 21--24 & F & Design student & 4 & 5--10 & Yes \\
\textbf{P8b} & 21--24 & F & Non-design student & 5 & 5--10 & Yes \\
\addlinespace[2pt]
\rowcolor{black!4}
\textbf{P9a} & 21--24 & M & Design student & 4 & 5--10 & Yes \\
\rowcolor{black!4}
\textbf{P9b} & 21--24 & M & Design student & 6 & 5--10 & Yes \\
\addlinespace[2pt]
\textbf{P10a} & 18--20 & M & Non-design student & 5 & $>$10 & Yes \\
\textbf{P10b} & 18--20 & M & Design student & 5 & $>$10 & Yes \\
\addlinespace[2pt]
\rowcolor{black!4}
\textbf{P11a} & 18--20 & F & Design student & 4 & $>$10 & Yes \\
\rowcolor{black!4}
\textbf{P11b} & 21--24 & F & Design student & 4 & 5--10 & Yes \\
\addlinespace[2pt]
\textbf{P12a} & 21--24 & M & Design student & 6 & $>$10 & Yes \\
\textbf{P12b} & 18--20 & F & Non-design student & 5 & 5--10 & Yes \\
\bottomrule
\end{tabular}

\end{table*}

\subsection{\duo{}: Shared Materials, Separate AI Chatbots}

\begin{figure*}[t]
\centering
\definecolor{interfaceShared}{HTML}{B95E62}
\definecolor{interfacePersonal}{HTML}{477AA8}
\includegraphics[width=\textwidth]{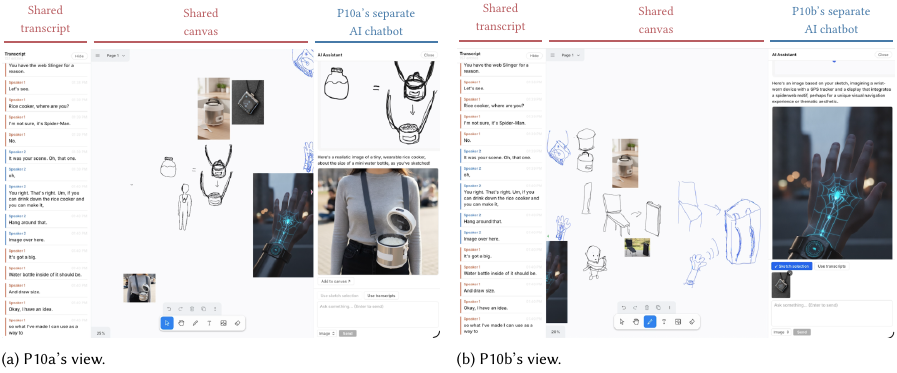}
\caption{\duo{} as used by P10a and P10b. Both collaborators access the same live transcript and shared canvas (red labels), while each uses a separate AI chatbot with its own interaction history (blue labels). The screenshots show different views of the shared canvas. Canvas selections and the accumulated transcript can be attached to an AI request, and generated images can be added to the canvas. Sharing these materials does not automatically synchronise the two AI histories.}
\Description{Original interface screenshots from P10a and P10b are displayed side by side. Above each, red rules identify the transcript on the left and shared canvas in the centre; a blue rule identifies the AI panel on the right. The canvas views show overlapping sketches and images at different positions and zoom levels. The two AI panels contain different generations and separate request histories.}
\label{fig:interface}
\end{figure*}

\duo{} combined a shared meeting workspace with separate AIs (Figure~\ref{fig:interface}). Its design drew on collaborative whiteboards such as Miro and the combination of discussion and shared materials in meeting platforms such as Microsoft Teams~\cite{miroWhiteboard,microsoftTeamsCollaboration}. Built with React and TypeScript, it provided three modules: Shared Canvas, Shared Transcript, and Separate Chatbot. The first two gave both collaborators access to the same visual materials and discussion record; the third allowed each to explore concepts with a separate AI. Participants could connect these spaces by submitting shared materials to their chatbot and bringing generated images back to the canvas.

\paragraph{Shared Canvas} The canvas provided a jointly editable space for sketches, text, and images, implemented with tldraw and its synchronisation library.\footnote{\url{https://tldraw.dev/}} Cloudflare Durable Objects synchronised room-level canvas state between devices, with canvas records persisted in SQLite and image assets stored in Cloudflare R2. Collaborator A used black ink and Collaborator B used blue ink to distinguish their drawing contributions. Both collaborators could edit the shared materials while navigating and zooming independently.

\paragraph{Shared Transcript} A continuously updated transcript made the pair's spoken discussion available for review and use as AI context. One client captured room audio and streamed it through a Node.js service to Google Cloud Speech-to-Text,\footnote{\url{https://cloud.google.com/speech-to-text}} using the \texttt{latest\_long} model with two-speaker diarisation. Final recognition results were broadcast to both devices as speaker-labelled entries in the shared panel.

\paragraph{Separate Chatbot} Each collaborator could request text suggestions or images through their own chat panel, using \texttt{gemini-2.5-flash} and \texttt{gemini-2.5-flash-image}, respectively, through the Gemini API.\footnote{\url{https://ai.google.dev/gemini-api/docs}} Each device maintained its room-specific AI history in browser-local IndexedDB storage. Requests reconstructed prior user prompts and attached images, together with AI-generated text and images, in chronological order before appending the current message, without history summarisation or retrieval. Text and image modes used the same history. Both chatbots received the same system instruction to support two people designing on a shared whiteboard, draw on supplied sketches, transcripts, and prompts, and respond in English in an open-ended, generative manner. The two AI histories remained separate.

\paragraph{Moving materials between modules} Alongside a written prompt, participants could attach canvas material through \emph{Use sketch selection} and discussion through \emph{Use transcripts}. The former captured selected canvas objects together as one PNG, including either collaborator's sketches and any generated images in the selection. The snapshot entered the request as image content and did not change with subsequent canvas edits. The latter attached all accumulated transcript text available at submission. Either or both could accompany the prompt, and the attachment controls reset after sending. In the other direction, \emph{Add to canvas} placed a generated image on the shared canvas while retaining it in the originating chat. The other collaborator could then select that image and submit it to their own chatbot. Thus, an image could travel from Collaborator A's chatbot to the shared canvas and then into Collaborator B's chatbot, alongside sketches or shared discussion. The speech transcript contained the pair's spoken discussion; AI outputs remained in the chat panels and on the canvas when added.

\subsection{User Study}
\label{sec:user-study}

\begin{figure}[htbp]
\centering

\includegraphics[width=0.5\textwidth]{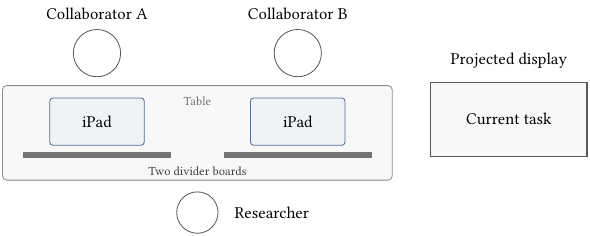}
\caption{In-person study setting (schematic plan view, not to scale). Collaborators A and B sat side by side, each using an iPad. The researcher sat opposite them, with two divider boards between the researcher and the collaborators. A projected display showed the current task; its position is indicative.}
\Description{A plan view shows two collaborators beside each other on one side of a table, an iPad in front of each, two divider boards across the middle, and a researcher seated on the opposite side. A projected task display is shown at the side of the meeting room.}
\label{fig:study-setting}
\end{figure}

Sessions took place in person in a large meeting room (Figure~\ref{fig:study-setting}). The two collaborators sat side by side at a table, each using an iPad to access \duo{}. One researcher sat opposite them. Two divider boards stood on the table between the researcher and the pair. A projected display showed the current task.

Participants were asked to develop concepts for a portable product to improve everyday commuting in Singapore. The brief allowed physical, digital, and service-oriented concepts. It provided a common context while leaving the problem and design direction open. Participants explored and refined concepts through discussion, sketches, and AI-generated representations.

Each session lasted approximately 85 minutes: a 10-minute introduction and tutorial, four 15-minute design phases, and a 15-minute joint post-session interview. Participants signed a consent form and completed a demographic questionnaire during the introduction. Table~\ref{tab:user-study} summarises the procedure and collected materials.

\begin{table*}[htbp]
\caption{User study procedure. The four design phases share the same data collection, shown in the merged cell below. AI use was optional throughout these phases.}
\label{tab:user-study}
\centering\small
\setlength{\tabcolsep}{4pt}
\renewcommand{\arraystretch}{1.2}
\begin{tabularx}{\textwidth}{@{}Q{0.11\textwidth}*{6}{Y}@{}}
\toprule
Stage & Introduction and tutorial & \cellcolor{Phase2!12}\textcolor{Phase2!65!black}{\textbf{Phase 1}}\par Individual Exploration & \cellcolor{Phase2!12}\textcolor{Phase2!65!black}{\textbf{Phase 2}}\par Team Synthesis & \cellcolor{Phase2!12}\textcolor{Phase2!65!black}{\textbf{Phase 3}}\par Team Adaptation & \cellcolor{Phase2!12}\textcolor{Phase2!65!black}{\textbf{Phase 4}}\par Team Consolidation & Post-session interview \\
\midrule
Time & 10 min & 15 min & 15 min & 15 min & 15 min & 15 min \\
\addlinespace[4pt]
Activities & Consent, demographic questionnaire, and system tutorial. & Develop two or three initial commuting concepts independently, with optional AI use. & Discuss and combine initial concepts into a shared design direction. & Develop the shared concept in response to an additional commuting requirement. & Organise representations and explain the final concept and rationale. & Revisit the process, AI use, contributions, difficulties, and desired support. \\
\midrule
Collected materials & Consent forms and demographic responses. & \multicolumn{4}{>{\raggedright\arraybackslash}p{\dimexpr(\textwidth-0.11\textwidth-12\tabcolsep)*4/6+6\tabcolsep\relax}}{\cellcolor{Phase2!12}\textbf{Across Phases 1--4:} Screen recordings, discussion transcripts, shared canvas materials, and AI interaction records (requests and outputs).} & Interview recording and reconstructed transcript. \\
\bottomrule
\end{tabularx}
\end{table*}

\paragraph{Introduction and tutorial (10 minutes)} Before the design task, participants signed an informed consent form and completed a demographic questionnaire. They then received a tutorial introducing the shared canvas, shared discussion transcript, and each collaborator's separate AI chatbot in \duo{}. The tutorial familiarised participants with the interface before they began the design activities. Operational assistance was available during the study.

\paragraph{Phase 1: Individual Exploration (15 minutes)} Each participant developed two or three initial concepts independently and could consult their own AI. This phase established the starting ideas, sketches, and AI-generated material that collaborators could subsequently bring into joint work. The task had an individual orientation within the shared workspace: collaborators could still speak and inspect each other's canvas activity.

\paragraph{Phase 2: Team Synthesis (15 minutes)} The collaborators discussed their initial concepts and decided what to select, relate, or combine into a shared design direction. They could use sketches, discussion, and either AI to explore the combination. Moving from individual generation to joint selection follows hybrid design processes~\cite{girotra2010generation}; this phase made the formation of shared understanding observable~\cite{clark1991grounding}.

\paragraph{Phase 3: Team Adaptation (15 minutes)} The researcher introduced an additional requirement concerning weather, place, transport mode, or pace. The pair reconsidered how its shared concept would address this requirement and developed it through further discussion, sketching, and optional AI use. The contextual change created an occasion to revisit both the problem and its possible solution~\cite{dorst2001coevolution}.

\paragraph{Phase 4: Team Consolidation (15 minutes)} The collaborators selected and organised representations of their concept and explained their design rationale. This activity brought together the decisions made during the preceding phases and made options, trade-offs, and remaining gaps explicit~\cite{maclean1991qoc}. Participants could continue sketching and consulting their AIs while preparing this account.

Throughout Phases 2--4, each pair worked towards a shared design concept while choosing when and how to use its separate AIs. Screen recordings captured the design activity alongside retained AI requests, outputs, and canvas materials. Figure~\ref{fig:p7-phases} in Appendix~\ref{sec:appendix-phase-example} illustrates P7's canvas across all four phases.

\paragraph{Post-session interview (15 minutes)} Immediately after the design activities, the researcher interviewed both collaborators together. They reviewed session recordings and revisited concept progression, AI use, transfers between separate AIs and the shared canvas, their respective contributions, and final decisions. The interview also asked about possible system improvements. We collected one interview per pair, yielding 12 joint interviews. The complete corpus is described in Appendix~\ref{sec:appendix-data-sources}, Table~\ref{tab:data-sources}; transcript preparation and qualitative coding are described below.

\subsection{Qualitative Data Analysis}
\label{sec:data-analysis}

\begin{figure*}[htbp]
\centering

\includegraphics[width=0.95\textwidth]{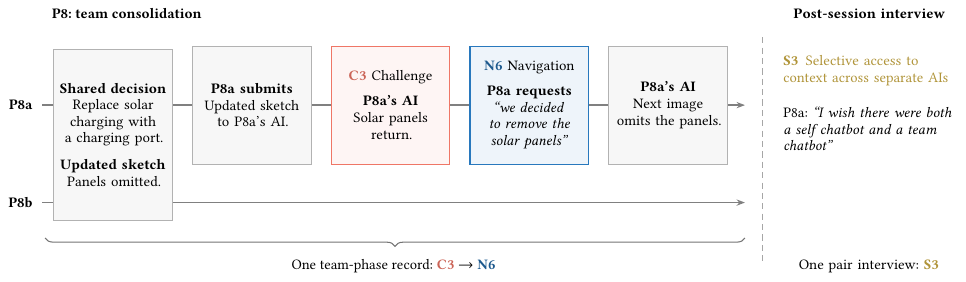}
\caption{\textbf{Example of assigning C, N, and S codes.} The AI's restoration of solar panels omitted from P8's updated sketch is coded as C3; P8a's explicit removal request is coded as N6. In the subsequent interview, P8a's suggestion for personal and team chats is coded as S3. C and N are assigned to the team-phase record, while S is counted once per pair interview. Quotations are excerpts; other text summarises the records.}
\Description{Two tracks show P8a and P8b's shared charging decision and P8a's subsequent AI requests: a red C3 output restores solar panels, and a blue N6 request removes them. A bracket marks one team-phase record. To the right, a separate interview column marked in yellow quotes P8a's post-session wish for both a personal and a team chatbot, illustrating a support code from a pair interview.}
\label{fig:analysis-example}
\end{figure*}

We used qualitative coding to analyse screen recordings, reconstructed transcripts of session discussions and post-session interviews, AI interaction records, and shared canvas materials. Transcripts were reconstructed from the recordings and manually checked for wording and speaker attribution.\footnote{Recordings were transcribed using Google Cloud Speech-to-Text Chirp v3: \url{https://docs.cloud.google.com/speech-to-text/v2/docs/chirp-model}.} The qualitative analysis followed concept development across all four phases, including Phase 1 to interpret starting concepts, design intentions, and the AI histories carried into collaboration. The unit for assigning and counting team-phase C and N codes was a complete team activity: 12 pairs across Phases 2--4 yielded 36 pair--phase records, connecting both collaborators' AI interactions with their discussion and sketching (Figure~\ref{fig:analysis-example}). To establish what an AI received, we examined the submitted text, selected images, attached live transcript, and preceding AI history. The reconstructed transcript used for analysis differs from the live transcript available for attachment during the session. For the focal interaction sequences, we traced each person's contribution, the resulting request, and the response. 

Three guiding questions frame the analysis: \textbf{C---Challenges:} What difficulties did pairs encounter when relating AI contributions to their shared design? \textbf{N---Navigation:} How did pairs navigate these difficulties and continue developing their shared design with AI? \textbf{S---Support:} What support did pairs suggest for developing shared design with AI? C and N concern observed encounters and responses; S concerns perceived support needs in the interviews.

We chose the pair--phase record to connect a difficulty with the discussion, sketch revisions, and AI requests through which collaborators addressed it. Each 15-minute design phase included drawing and discussion as well as AI use. Across the 36 team-phase records, the two collaborators together submitted 141 AI requests: a mean of 3.92 per record (median 3; range 0--13), including records with no requests. We counted submitted user messages in both AI logs, using their phase assignments in the chronological records; individual exploration, tutorial, and post-task interview requests were excluded. This unit allowed us to examine how a challenge developed across representations and requests within a shared task, while retaining request-level evidence and response timing. It provides a coarser account than turn-level coding and does not estimate the frequency of individual difficulties within a phase.

 Alongside pair--phase counts, we report the number of distinct pairs assigned each C or N code, counting a pair once per code across all team phases. A pair can contribute to multiple codes. Counts for each navigation direction use the union of pairs with any code in that direction; pairs using both directions are counted in each. These pair counts describe the spread of challenge-linked encounters across the 12 pairs, while record counts capture their occurrence across phases.

\begin{figure*}[htbp]
\centering

\includegraphics[width=0.8\textwidth]{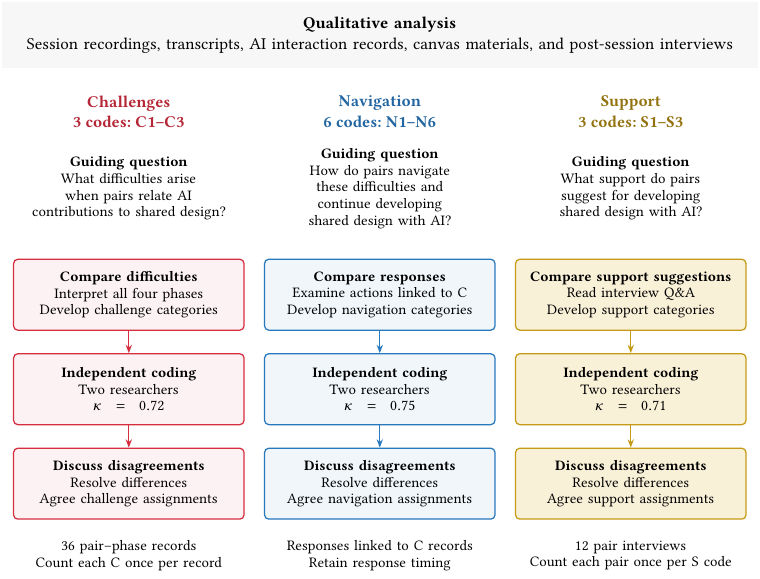}
\caption{Qualitative coding of challenges (C1--C3, red), navigation (N1--N6, blue), and support (S1--S3, yellow). Each column shows category development, independent coding, and resolution of disagreements. C and N concern task records and responses, while S concerns suggestions in pair interviews.}
\Description{Three parallel columns show three challenge codes C1 through C3 in red, six navigation codes N1 through N6 in blue, and three support codes S1 through S3 in yellow. Each column includes a guiding question, category development, independent coding, and resolution of disagreements. Kappa values are 0.72 for challenges, 0.75 for navigation, and 0.71 for support.}
\label{fig:analysis-workflow}
\end{figure*}

Two researchers developed three challenge (C) categories by reviewing full sessions chronologically, connecting participants' reflections to interaction sequences, and comparing difficulties within and across pairs (Table~\ref{tab:challenges}). Coding required evidence of an unresolved question or trade-off about an AI proposal (C1), a shared meaning missing from a response (C2), or an agreed update missing from a later output (C3). For C3, we traced the earlier design state, the pair's revision, and the subsequent mismatch. Multiple C codes could be assigned to a record, in the phase where the difficulty became evident. Two researchers independently applied the challenge codes ($\kappa = 0.72$) and resolved disagreements through discussion. Figure~\ref{fig:analysis-workflow} summarises the analysis.

For navigation (N), the researchers compared responses to the identified challenges and developed six categories (Table~\ref{tab:navigation-codebook}). N1--N3 concern reworking AI proposals, developing their uses, and assessing their fit with shared requirements. N4--N6 concern bringing shared materials to AI, explaining shared meanings, and communicating decisions. Responses were linked to the challenge they addressed, including actions in subsequent phases. Each strategy was counted once per challenge-labelled record, with multiple N codes assigned when distinct actions occurred. Two researchers independently applied the navigation codes ($\kappa = 0.75$) and resolved disagreements through discussion.

C and N identify a difficulty and the action taken in relation to it. A spoken assessment could both articulate an unresolved fit (C1) and assess a proposal against a requirement (N3); a question alone did not warrant N3. Appendix~\ref{sec:appendix-codebooks} illustrates this boundary and how challenge-linked coding treats actions that cross directions. We interpret the six strategies as bridging work, organised by direction and by the content involved: materials and form, meaning and use, and decisions and criteria. Tables~\ref{tab:challenge-overview} and~\ref{tab:navigation-counts} report record counts and distinct-pair counts.

We analysed the post-session interviews for three perceived support codes (S): jointly assessing and revisiting AI proposals (S1), jointly expressing shared design intentions to AI (S2), and selective access to context across separate AIs (S3). Each answer was read with its preceding question and related session sequence to identify concrete suggestions for support. Suggestions could receive multiple S codes. Two researchers independently applied the codes ($\kappa = 0.71$) and resolved disagreements through discussion. Each pair counted once per S code when its interview contained at least one relevant suggestion. Table~\ref{tab:support-codebook} defines these codes; Section~\ref{sec:interview-support} presents the accounts.

\FloatBarrier
\section{Results}
\label{sec:results}

Pairs encountered challenges in judging AI proposals, expressing shared meanings, and carrying agreed changes into later responses (Section~\ref{sec:challenges}). They navigated these challenges by developing AI proposals together and communicating shared design to AI (Section~\ref{sec:navigation}). Their interview suggestions concerned support for joint judgement, joint expression, and context sharing (Section~\ref{sec:interview-support}).

The analysis covered 36 pair--phase records. Twenty records across ten pairs received at least one challenge code; P8's Phase 4 received both C2 and C3, yielding 21 assignments. Sixteen records received no C code. Tables~\ref{tab:challenge-overview} and~\ref{tab:navigation-counts} report the challenge and navigation distributions, with detailed definitions in Appendix~\ref{sec:appendix-codebooks}.

\subsection{Challenges Pairs Encounter in Connecting Shared Design with AI}
\label{sec:challenges}

The three challenges concern how pairs connect a developing design with AI. C1 arises when the fit between a proposal and shared purposes remains unresolved. In C2, AI responses do not express a meaning the pair has established. In C3, later responses miss an agreed design change. Together, they concern forming shared purposes, expressing shared meanings, and carrying decisions forward.

\begin{table*}[htbp]
\centering\small\setlength{\tabcolsep}{5pt}\renewcommand{\arraystretch}{1.15}
\caption{Challenges across 36 pair--phase records. The Pairs column counts distinct pairs per code, with each pair counted once across phases; Pair IDs identifies them. Pair counts overlap across codes. P8's Phase 4 has both C2 and C3 and counts once in the Any C row.}
\label{tab:challenge-overview}\label{tab:challenge-counts}
\begin{tabularx}{\textwidth}{@{}l Y rrrrr Q{0.18\textwidth}@{}}
\toprule Code & Challenge & \phasebadge{2} & \phasebadge{3} & \phasebadge{4} & Records & Pairs & Pair IDs \\ \midrule
\ccode{C1} & Unresolved Fit between AI Proposals and Shared Design Purposes & \cellcolor{C1!8}2 & \cellcolor{C1!8}4 & 0 & \cellcolor{C1!8}6 & 5 & P1, P2, P8, P9, P11 \\
\ccode{C2} & Shared Design Meanings Not Expressed in AI Responses & \cellcolor{C2!8}4 & \cellcolor{C2!8}3 & \cellcolor{C2!8}6 & \cellcolor{C2!8}13 & 8 & P1, P3, P5, P7, P8, P9, P10, P12 \\
\ccode{C3} & Agreed Design Changes Missing from Later AI Responses & 0 & 0 & \cellcolor{C3!8}2 & \cellcolor{C3!8}2 & 2 & P8, P11 \\ \midrule
Any C & Records with at least one challenge & 6 & 7 & 7 & 20 & 10 & P1--P3, P5, P7--P12 \\
\bottomrule\end{tabularx}
\end{table*}

C2 was the most widespread challenge, appearing in thirteen records across eight pairs. Four pairs (P3, P9, P10, P12) encountered it in multiple phases as their concept developed. C1 appeared in Synthesis and Adaptation, while both C3 encounters occurred in Consolidation. The following cases show how these challenges arose within the study activities.

\subsubsection{C1: Unresolved Fit between AI Proposals and Shared Design Purposes}
\label{sec:challenge-open}

C1 involved five pairs (P1, P2, P8, P9, P11) across six records in Team Synthesis and Team Adaptation (Phases 2--3). While combining initial ideas or adapting a concept to a new commuting situation, pairs invited AI to propose connections, features, and uses. They then needed to work out how these proposals served their shared purposes and constraints. The unresolved fit concerned a reinterpreted purpose, a feature whose use needed development, or a trade-off introduced by the proposal.

\begin{figure*}[htbp]
\centering
\includegraphics[width=\textwidth]{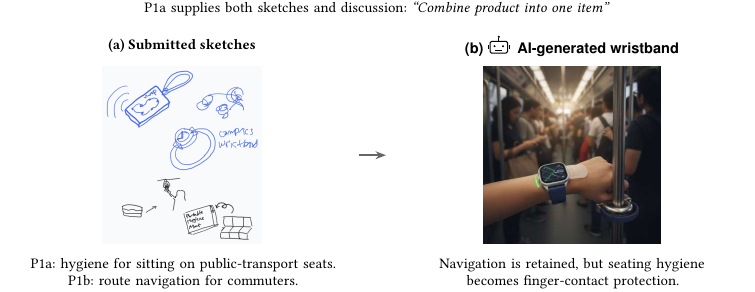}
\caption{\textbf{C1 example: a combined proposal changes what a feature is for.} Both collaborators' sketches and discussion are supplied to P1a's AI. Its wristband proposal retains navigation but recasts seating hygiene as finger-contact protection. The depiction follows P1a's selection of the wristband proposal. During Phase 3, P1b questions the need for finger protection and P1a finds it inconvenient. The images are study artefacts; the combination request is quoted.}
\Description{Both starting sketches appear beside the AI-generated wristband. The sketch includes a seating-hygiene mat; the wristband depicts finger-contact protection, whose practical fit the pair later questions.}
\label{fig:c1-evidence}
\end{figure*}

A combination could reinterpret what a starting concept was meant to accomplish. P1a brought seating-hygiene and handle-extension concepts; P1b brought route-navigation concepts. P1a supplied both collaborators' sketches and live discussion, requesting \dquote{Combine product into one item}. The request drew on P1a's earlier AI explorations of hygiene, seat sensing, and reach support. The AI proposed a mat and a wristband. The mat retained seating hygiene and a standing grip, and repurposed its display for seating information. The wristband retained navigation while recasting seating hygiene as finger protection for touching buttons and poles. In Phase 3, P1b questioned the need for finger protection beyond bus use, and P1a found it inconvenient. Figure~\ref{fig:c1-evidence} shows this encounter.

A proposed feature could also leave its role open to joint consideration. P11b wanted a countdown to help decide whether to hurry for a train. In response to a combination request, P11a's AI proposed a pouch with an e-ink countdown display and a phone holder, describing use while seated or waiting. P11b found a screen on the pouch odd, while P11a questioned whether simply attaching a phone integrated their concepts. Their subsequent discussion developed a condition for using the display: it should appear near a platform (Figure~\ref{fig:navigation-ai-human}). The countdown itself came from the earlier concept and AI proposal; the pair developed its situational use.

Other additions raised trade-offs against shared requirements. For P8, proposed storage and further bottle functions prompted questions about collapsibility, weight, and power. An attractive addition could change the implications of the concept for both collaborators. Across these cases, pairs jointly considered how an AI proposal would contribute to their design.

\subsubsection{C2: Shared Design Meanings Not Expressed in AI Responses}
\label{sec:challenge-meaning}

C2 involved eight pairs across thirteen records in Team Synthesis, Adaptation, and Consolidation (Phases 2--4). Pairs established shared meanings through discussion, pointing, and sketching, including where components belonged, how a service covered several locations, and how someone would use a concept. Conveying those meanings to AI required selecting materials and formulating requests. Responses sometimes omitted or depicted these relationships differently, even when recognisable components or explicit wording were supplied.

\begin{figure*}[htbp]
\centering
\includegraphics[width=0.85\textwidth]{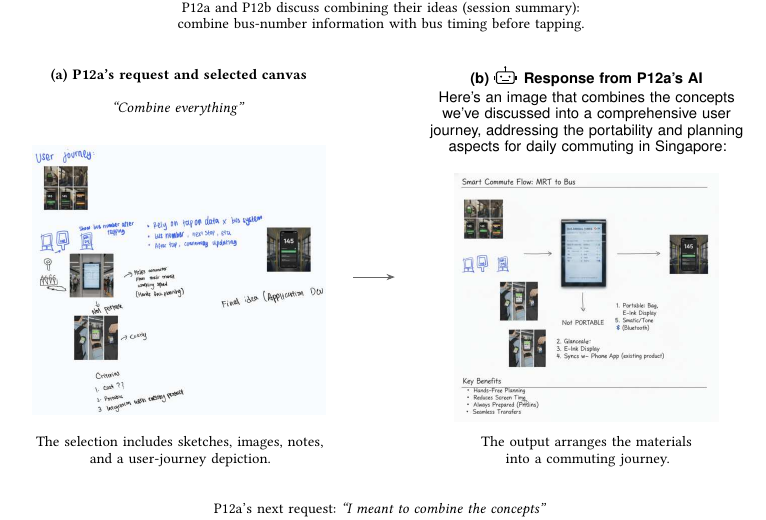}
\caption{\textbf{C2 example: the pair discusses a functional combination, but the AI request does not specify it.} P12 discusses combining bus-number and bus-timing information. P12a submits selected canvas materials with \dquote{Combine everything}; the AI instead arranges them as a user journey. P12a then clarifies \dquote{I meant to combine the concepts}. The discussion is summarised from the session transcript; the request, clarification, and AI excerpt are verbatim. The canvas selection and output are original study images.}
\Description{A summary of P12a and P12b's discussion appears above two panels. The first shows the canvas material P12a submitted with the request Combine everything; the second shows the AI's commuting-journey image and a verbatim response excerpt. Below, P12a clarifies that the request was to combine the concepts.}
\label{fig:c2-combination-evidence}
\end{figure*}

An intended combination could remain implicit in the request. In Team Synthesis, P12 discussed integrating their ideas: P12b proposed combining bus-number information with bus timing before tapping. P12a then selected canvas materials and asked their AI to \dquote{Combine everything}. The selection contained several sketches, generated images, and notes, including a user-journey depiction, but the request did not specify the functional relationship discussed by the pair. The AI arranged the material into a commuting journey, describing it as a \dquote{comprehensive user journey}. P12a corrected this interpretation: \dquote{I meant to combine the concepts}. Here, the materials reached one AI, while the intended operation on those materials remained underspecified. Figure~\ref{fig:c2-combination-evidence} shows the discussion, submitted material, and response.

\begin{figure*}[htbp]
\centering
\includegraphics[width=0.85\textwidth]{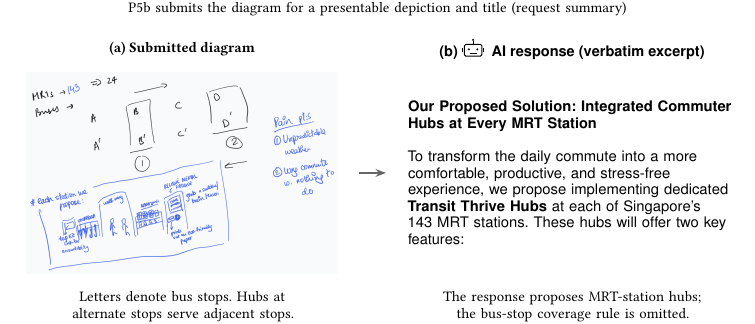}
\caption{\textbf{C2 example: the shared diagram reaches the AI, but its operating rule is missing from the response.} P5b submits a diagram of lettered bus stops and hub locations. The AI describes MRT-station hubs without the intended bus-stop coverage rule. The diagram is a study artefact. The request is summarised; the response panel reproduces a verbatim excerpt from the AI log in a sans-serif font.}
\Description{The original lettered hub diagram is shown beside a verbatim excerpt of the AI response, with a robot icon identifying the AI panel. The response describes MRT-station services but omits the rule that alternate hubs serve adjacent bus stops.}
\label{fig:c2-evidence}
\end{figure*}

A shared diagram supplied to an AI did not always convey its operating rules. Before the request, P5a proposed locating a hub at alternate bus stops so that it covered neighbouring stops in either direction; P5b replied, \dquote{Okay, that sounds like a plan. I'll draw a sketch of one of these stations}. This exchange connects the coverage rule to the pair's subsequent sketching. P5b submitted the diagram for a presentable depiction and title, with no live transcript attached and no written explanation of the lettered stops or coverage rule. The response described integrated hubs at MRT stations without the bus-stop coverage rule. The diagram supplied components, but the output omitted the relationship that explained how the concept worked. Figure~\ref{fig:c2-evidence} shows this encounter.

A requested interaction could likewise be depicted incorrectly despite recognisable components. In Phase 3, P12 repeatedly specified a physical EZ-Link card connected to a phone app, distinguishing it from a card on the phone. The participant later identified the continuing phone-based depiction as a mismatch with this intention. Across these cases, pairs needed to explain how the depicted components related to the design they had discussed.

\subsubsection{C3: Agreed Design Changes Missing from Later AI Responses}
\label{sec:challenge-current}

C3 arose in two pairs, P8 and P11, during Team Consolidation (Phase 4). As they prepared to present their design, the pairs selected features and revised sketches using developments from discussion, canvas work, and AI exchanges. Later responses missed some of these agreed changes. In P11, an output omitted the combined concept developed through the other AI; in P8, an output restored a rejected feature despite receiving an updated sketch.

One form was the omission of developments made through the other AI. P11b initially explored a countdown app, while P11a explored a folding chair. The pair combined them by placing a screen on the chair's pouch, then developed adjustable height and a detachable screen through P11a's AI. When P11b later asked their AI to illustrate automatic countdown near train platforms, the output depicted a phone and platform display. The request named the tablet and proximity function but included no image or live transcript; P11b's preceding AI history concerned the original app. The countdown function remained, but the combined chair-and-pouch concept developed through P11a's AI was absent. P11a noticed the missing transcript, whereas P11b initially considered the wording descriptive enough. P11b attached the live transcript in subsequent requests (N4); the next reply then referred to a portable product with an integrated screen.

\begin{figure*}[htbp]
\centering
\includegraphics[width=\textwidth]{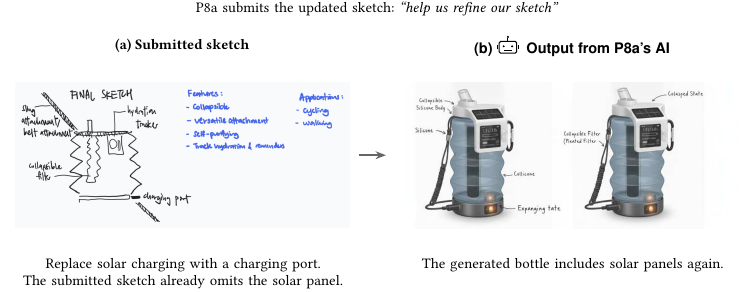}
\caption{\textbf{C3 example: an updated sketch does not prevent a rejected feature from returning.} P8's submitted sketch already omits solar panels and labels a charging port, but P8a's AI restores the panels. The images are the study input and output; the request is quoted, and the decision is summarised from the record.}
\Description{The shared bottle sketch without solar panels is shown beside the AI output with solar panels. The submitted sketch includes a charging port.}
\label{fig:c3-evidence}
\end{figure*}

Another form was the return of a rejected feature despite an updated sketch. Earlier, P8a had asked AI to include solar-powered filtration. During Consolidation, the pair reconsidered which features to retain:
\begin{quote}\small
\textbf{P8b:} \dquote{I think we can remove the solar panel and the additional storage compartment, for the weight constraint.}\\
\textbf{P8a:} \dquote{Okay --- but if we remove the solar panel, then we won't have the purification anymore.}\\
\textbf{P8b:} \dquote{Charge your bottle [instead].}
\end{quote}
This exchange links the proposed removal to the need to preserve purification. The subsequent shared sketch omitted solar panels and labelled a charging port, and P8a's final presentation confirmed this replacement. P8a then submitted that sketch with \dquote{help us refine our sketch}, attaching no live transcript and giving no explicit removal instruction. The AI restored the panels. P8a then stated the removal decision (N6), and the next image omitted them. Figure~\ref{fig:c3-evidence} shows the input and mismatching output.

C3 concerns carrying a revision forward as it supersedes an earlier design state. P11's later request did not supply the combined concept. P8 supplied a current drawing, but the rejected solar-based design remained in the AI history. Both pairs needed to make their updated design explicit in a subsequent request.

These encounters connect AI responses to different aspects of the pair's design: what a proposal is for, how its components work together, and which changes should carry forward. The next section examines how pairs navigated these challenges.

\subsection{How Pairs Bridge Shared Design and AI Interaction}
\label{sec:navigation}

Pairs navigated the challenges in two directions. Five pairs worked AI proposals into shared design (N1--N3), addressing questions about their fit during Synthesis and Adaptation. Nine pairs brought shared materials, meanings, and decisions to AI (N4--N6), addressing expression and revision across the team activities. Four pairs (P1, P8, P9, P11) used both directions. Table~\ref{tab:navigation-counts} shows the strategies linked to each challenge.

\begin{table*}[htbp]
\centering\small\setlength{\tabcolsep}{3pt}\renewcommand{\arraystretch}{1.15}
\caption{How pairs bridge shared design and AI interaction. Counts refer to challenge-linked pair--phase records. The Records column counts each record once across C categories; P8's Phase 4 links N6 to both C2 and C3. Later responses remain linked to the challenge they address. The Pairs column counts distinct pairs per code, with each pair counted once across phases; Pair IDs identifies them. Pair counts overlap across codes.}
\label{tab:navigation-counts}
\begin{tabularx}{\textwidth}{@{}l Y rrrrr l@{}}
\toprule Code & Navigation & \ccode{C1} & \ccode{C2} & \ccode{C3} & Records & Pairs & Pair IDs \\ \midrule
\multicolumn{8}{@{}l}{\textit{AI $\rightarrow$ Shared Design}} \\ \addlinespace[3pt]
\nshort{1} & Reworking AI Proposals for Shared Design & \cellcolor{N1!8}4 & 0 & 0 & \cellcolor{N1!8}4 & 3 & P1, P2, P8 \\
\nshort{2} & Developing Uses for AI Proposals & \cellcolor{N2!8}2 & 0 & 0 & \cellcolor{N2!8}2 & 2 & P9, P11 \\
\nshort{3} & Assessing AI Proposals against Shared Requirements & \cellcolor{N3!8}5 & 0 & 0 & \cellcolor{N3!8}5 & 4 & P1, P8, P9, P11 \\
\midrule\multicolumn{8}{@{}l}{\textit{Shared Design $\rightarrow$ AI}} \\ \addlinespace[3pt]
\nshort{4} & Bringing Shared Design Materials to AI & 0 & \cellcolor{N4!8}2 & \cellcolor{N4!8}1 & \cellcolor{N4!8}3 & 3 & P3, P10, P11 \\
\nshort{5} & Explaining Shared Design Meanings to AI & 0 & \cellcolor{N5!8}7 & 0 & \cellcolor{N5!8}7 & 5 & P3, P5, P7, P9, P12 \\
\nshort{6} & Communicating Shared Design Decisions to AI & 0 & \cellcolor{N6!8}5 & \cellcolor{N6!8}1 & \cellcolor{N6!8}5 & 5 & P1, P3, P8, P9, P10 \\
\bottomrule\end{tabularx}
\end{table*}

\subsubsection{\texorpdfstring{AI $\rightarrow$ Shared Design}{AI to Shared Design}}
\label{sec:navigation-from-ai}

Pairs revised proposals, developed their uses, and assessed their fit with shared requirements. These actions could lead to a new feature or a reason to retain, adapt, or reject an AI suggestion. In this direction, navigation contributed to developing the shared design itself.

\paragraph{N1: Reworking AI Proposals for Shared Design} Three pairs treated AI proposals as revisable material, preserving a broader design direction while changing features that did not serve it. P1 retained interest in combining hygiene and navigation, but replaced the questioned finger-protection feature with a folding clip through discussion and sketching. P2 developed a cap-and-slot attachment instead of the proposed silicone strap. P8 revised bottle variants and later removed features after discussing weight and power. Through these revisions, pairs developed alternatives to the features they had questioned.

\begin{figure*}[htbp]
\centering
\includegraphics[width=\textwidth]{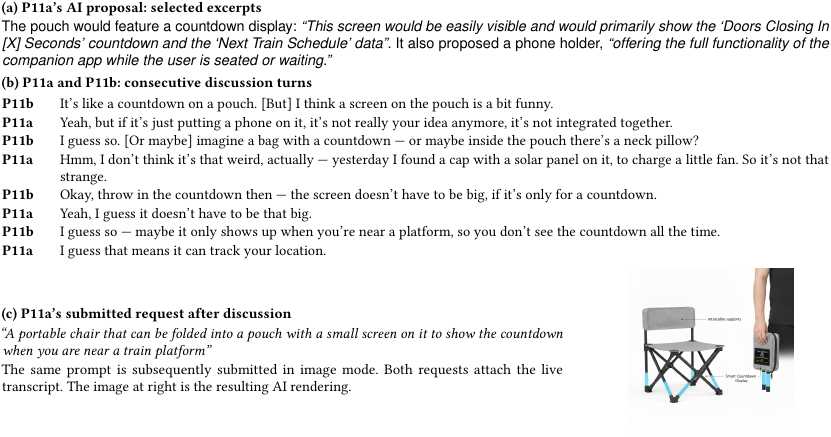}
\caption{\textbf{N2 and N3 example: judging and developing an AI-proposed feature.} Following P11's C1 encounter described in Section~\ref{sec:challenge-open}, the pair questions whether attaching a phone integrates the two concepts (N3), then establishes a proximity-triggered countdown role (N2). The AI excerpts and submitted request are from the logs; the discussion reproduces consecutive turns from the reconstructed transcript. The AI already proposed a countdown display; the pair developed its proximity condition. The final panel shows the AI rendering after the same request was submitted in image mode. The top excerpts use sans-serif text to identify AI-produced material.}
\Description{Excerpts of the initial AI proposal show a countdown display and phone holder. Eight consecutive participant turns show the pair questioning the combination and developing the display's proximity condition. The submitted request and an AI rendering show this development carried into the next interaction.}
\label{fig:navigation-ai-human}
\end{figure*}

\paragraph{N2: Developing Uses for AI Proposals} Two pairs developed how proposed features would be used. P11's discussion moved from questioning the combination to specifying when its screen should display the countdown (Figure~\ref{fig:navigation-ai-human}). P11b proposed limiting the display to moments near a platform; P11a acknowledged the resulting need to track location and included the proximity condition in the next request. Their discussion established where the display would be used and the location tracking this would require. P9 explored how a proposed shape-memory-alloy mechanism could open the shoe while considering its cost. In both cases, the pair supplied an account of operation or use that the proposal had left unresolved. Developing a use helped the pair establish how a proposed feature could belong in the shared concept.

\paragraph{N3: Assessing AI Proposals against Shared Requirements} Four pairs brought proposals into joint judgement, making their implications for shared purposes explicit. Assessment could support an existing preference, expose a trade-off, or motivate further development. P8 already preferred the bottle when AI advice about cycling reinforced concerns about a wearable umbrella. Later, suggestions from separate AI histories entered a shared discussion of the bottle's weight and power demands. Alongside P1's convenience concern, P11's integration question, and P9's cost discussion, this shows how judgement situated AI suggestions within the pair's developing requirements.

\subsubsection{\texorpdfstring{Shared Design $\rightarrow$ AI}{Shared Design to AI}}
\label{sec:navigation-to-ai}

Pairs brought shared materials to AI (N4), explained shared meanings (N5), and communicated design decisions (N6). These strategies addressed what AI needed to work with: a relevant reference, an explanation of a relationship or use, or a statement of the intended change.

\paragraph{N4: Bringing Shared Design Materials to AI} Across P3, P10, and P11, collaborators used material already available in shared work to provide a useful basis for the next AI response. P10 and P11 supplied discussion to situate sketches or features within the developing concept; P3 borrowed a representation generated through the other AI. Pairs selected materials that connected the request to their developing design.

\begin{figure*}[htbp]
\centering
\includegraphics[width=\textwidth]{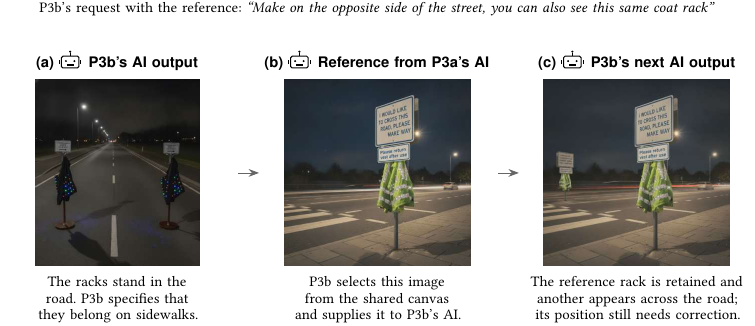}
\caption{\textbf{N4 example: supplying a reference to express a shared spatial intention.} P3b follows a sidewalk clarification with an image generated through P3a's AI. The transfer occurs in Phase 3 and addresses a C2 rack-placement challenge first observed in Phase 2: the pair wanted sidewalk racks, but P3b's AI placed them in the roadway. The next output uses the reference but still requires placement correction. Images are original study outputs; the request is quoted from the log. Robot icons identify AI-produced panels.}
\Description{Three original AI images show racks in the road, the partner's sidewalk-rack reference, and the next output incorporating that reference. The collaborator supplies the reference and a request to AI.}
\label{fig:navigation-human-ai}
\end{figure*}

P3 reused a generated reference to clarify rack placement. After repeated difficulty obtaining sidewalk racks from P3b's AI, P3b selected P3a's generated scene from the shared canvas and requested, \dquote{Make on the opposite side of the street, you can also see this same coat rack}. The next output used the sidewalk reference, although the far rack still needed repositioning (Figure~\ref{fig:navigation-human-ai}). Reusing the generated reference provided a basis for clarifying the remaining placement issue. The Phase 3 transfer remained linked to the difficulty first observed in Phase 2.

\begin{figure*}[htbp]
\centering
\includegraphics[width=0.85\textwidth]{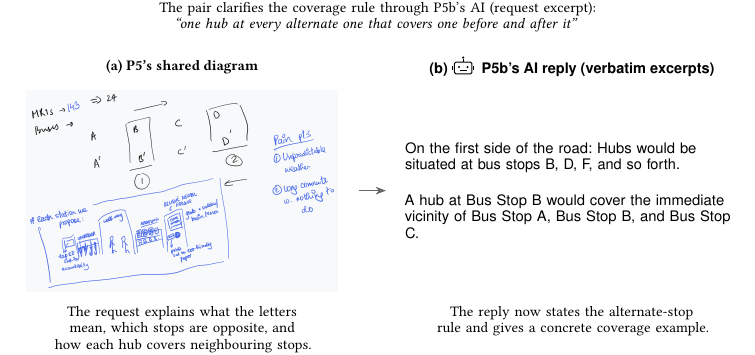}
\caption{\textbf{N5 example: explaining the operating rule of a shared diagram.} The clarification submitted through P5b's AI explains the lettered stops, opposite-side stops, and alternate-hub coverage. The next textual reply states the rule and gives the example of hub B serving A, B, and C. The diagram is an original study artefact; the request and reply are verbatim excerpts, with Markdown emphasis omitted from the reply.}
\Description{P5's shared diagram of lettered bus stops appears beside excerpts of the next AI reply. The request explains that each alternate hub covers the stops before and after it. The reply specifies hubs at B, D, and F and states that hub B covers A, B, and C.}
\label{fig:navigation-n5}
\end{figure*}

\paragraph{N5: Explaining Shared Design Meanings to AI} Five pairs clarified meaning across seven records by explaining an intended object, relationship, or operating rule. P5 had agreed that hubs at alternate bus stops would serve neighbouring stops on both sides of the road. After the AI's response omitted this coverage rule, the pair made the relationship explicit in a clarification submitted through P5b's AI. The request explained what the letters represented, which stops were opposite one another, and how each hub covered adjacent stops: \dquote{one hub at every alternate one that covers one before and after it}. The next textual reply stated that hub B covered stops A, B, and C (Figure~\ref{fig:navigation-n5}). This clarification supplied the operating rule needed to interpret the shared diagram.

\paragraph{N6: Communicating Shared Design Decisions to AI} Five pairs explicitly stated a required change, turning a shared decision into an instruction for AI. P8 assessed weight and power concerns (C1), revised the bottle concept, and removed solar panels from the shared sketch. When a later AI output restored the panels (C3), P8a requested, \dquote{we decided to remove the solar panels, can you update the sketch, could u also do a belt attachment?} The next image omitted them. P8 thus both revised an AI proposal together and communicated an agreed change to AI. P10 similarly requested a smaller cooker after judging its depicted size excessive. These requests specified how the next output should reflect the pair's decision. Across N4--N6, collaborators made different aspects of shared design explicit to AI by supplying a relevant reference (N4), explaining what it meant or how the concept worked (N5), or stating what should change (N6).

Taken together, the six navigation strategies show two directions of work: pairs reworked, developed, and assessed AI proposals within shared design, and supplied materials, explained meanings, and communicated decisions to AI. Four pairs used both directions during the study.

\subsection{Support Pairs Sought for Shared Design with AI}
\label{sec:interview-support}

Pairs sought support for jointly assessing and revisiting AI proposals (S1), contributing to the expression of shared intentions (S2), and accessing relevant context across separate AIs (S3). Table~\ref{tab:support-results} reports these suggestions across the twelve interviews; Appendix~\ref{sec:appendix-codebooks} defines the codes. The accounts below connect these support needs to the observed challenges and navigation.

\begin{table*}[htbp]
\centering\small\setlength{\tabcolsep}{5pt}\renewcommand{\arraystretch}{1.18}
\caption{Perceived support needs across 12 pair interviews. Each pair counts once per S code; codes may overlap and do not imply agreement within a pair.}
\label{tab:support-results}
\begin{tabularx}{\textwidth}{@{}l Y r Q{0.25\textwidth}@{}}
\toprule Code & Perceived support need & Count & Pair interviews \\ \midrule
\scode{1} & Jointly assessing and revisiting AI proposals & \cellcolor{S1!8}3 & P6, P9, P10 \\ \addlinespace[3pt]
\scode{2} & Jointly expressing shared design intentions to AI & \cellcolor{S2!8}3 & P3, P4, P6 \\ \addlinespace[3pt]
\scode{3} & Selective access to context across separate AIs & \cellcolor{S3!18}7 & P3, P5, P6, P7, P8, P10, P11 \\
\bottomrule\end{tabularx}
\end{table*}

\paragraph{S1: Jointly assessing and revisiting AI proposals} Three pair interviews (P6, P9, P10) sought help with evaluating or reopening design directions. This connects to C1: a proposal can look coherent while its purpose, use, or suitability still needs joint consideration. N1--N3 show collaborators doing that work themselves; the interviews identify resources they wanted for continuing it. P6b proposed questions about a scenario's purpose and limitations, while a P9 participant wanted studies and numbers to assess a concern grounded in personal experience. P10b's proposal to revisit an earlier point in a shared timeline addressed another part of the same need: being able to revise a direction and its AI context after reconsideration. P2b's concern about becoming \dquote{fixed on the thing that's been generated} helps explain the value of reopening alternatives, although this reflection was not counted as a concrete suggestion. S1 concerns retaining room and grounds for shared judgement as AI proposals become more concrete.

\paragraph{S2: Jointly expressing shared design intentions to AI} Three pair interviews (P3, P4, P6) sought a more direct role in clarification. P3 exposes a limit of reference sharing: one participant could take the other's image into their AI, but additions still had to be relayed verbally. The other participant considered canvas images sufficient and placed less value on sharing text history. Their differing views distinguish seeing the same material from being able to contribute to the request that uses it. This extends the C2 finding: even when both people understand a concept, its expression may depend on one collaborator expressing it to AI through N5 or N6. P4a's proposal that both people circle regions of the same image would let either collaborator indicate the intended change directly; P6b's wish that \dquote{the AI could also ask questions back} would make gaps in expression available for clarification. S2 therefore concerns participation in conveying shared meaning, beyond the access to context captured by S3.

\paragraph{S3: Selective access to context across separate AIs} Seven pair interviews (P3, P5, P6, P7, P8, P10, P11) sought ways to make relevant AI context available to shared work. P8a's account makes the coordination issue concrete: \dquote{we could have just done it in one, as we were both trying to refine the final design}. Yet P8a wanted a team chat alongside personal chats, and P5b wanted to share \dquote{part of my history with her directly}. These accounts point to choices about which material and activities become shared. This relates to C3, where later AI outputs did not reflect shared developments, and to the manual transfer observed in N4. It also supports the reference reuse seen with C2. P10b's request to identify who generated a response and P11's proposal to share summaries or rationales further indicate that the source and relevance of transferred material mattered.

The suggestions concern both participation and context: P3 wanted to contribute additions directly, P8 wanted a team chat alongside personal chats, and P10 wanted to revisit an earlier direction. The following implications develop these requests into support for joint judgement, joint expression, and communication of evolving design context.

\section{Design Implications}
\label{sec:design-implications}

Pairs navigated challenges by developing AI proposals together and communicating shared design to AI. We derive three implications from these encounters and participants' suggestions for support. Each connects an observed difficulty to the tasks and interaction conditions in \duo{}, then considers how mechanisms from prior systems could be adapted to support future collaboration. These are design opportunities grounded in an exploratory study; their effects remain to be evaluated.

\subsection{Supporting Joint Judgement as Design Purposes Evolve}

Five pairs encountered unresolved fit between AI proposals and shared design purposes (C1). Their responses included revising features, developing uses, and assessing proposals against shared requirements (N1--N3). P1 replaced the questioned finger-protection feature, P11 developed when a countdown screen should activate, and P8 considered additions against weight and power constraints. These encounters show that a proposal's contribution to the design depended on further joint consideration. Participants' requests for questions about purpose, supporting evidence, and ways to revisit earlier directions (S1) identify resources for this work.

The study tasks help explain why these questions became salient. Synthesis asked pairs to combine independently developed concepts, while Adaptation introduced an additional commuting requirement. In both activities, a proposed combination or feature had to be considered in relation to a developing use situation. \duo{} let pairs generate proposals and share their depictions, while discussion and sketching provided the means to examine their implications. It offered no dedicated support for comparing proposals against the pair's purposes or recording unresolved trade-offs. We interpret C1 within this combination of open-ended tasks and generative support: pairs needed to work out the relevance of a proposal as they developed the concept. This interpretation is consistent with accounts of reflection-in-action and problem--solution co-evolution~\cite{schon1983reflective,dorst2001coevolution}; the phase pattern reflects the activities elicited by our procedure.

Prior systems suggest ways to make this consideration more explicit. IdeaSynth represents research ideas as editable facets, supporting alternatives and refinement through literature-grounded feedback~\cite{pu2025ideasynth}. Botender uses generated interaction scenarios to prompt community discussion of desired bot behaviour and reveal disagreements~\cite{kuo2026botender}. Although these systems address research ideation and community bots, they offer two mechanisms relevant here: keeping aspects of an idea open to revision, and using concrete situations to examine what a proposal would do. For pairs working with visual concepts, these mechanisms could support discussion of a feature's purpose and use alongside its depiction.

Future systems could let either collaborator attach an intended use, an unresolved question, or a design criterion to a selected proposal. Pairs could request alternative features or short use scenarios linked to those annotations, then compare their consequences together. For a proposal like P11's screen, alternative scenarios could help the pair discuss when the information would be useful and what activation would require. Earlier alternatives and the reasons for retaining or rejecting them should remain accessible. DesignTrace offers a precedent for tracking visual alternatives through branching histories~\cite{peng2026designtrace}; linking those branches to the pair's purposes would extend this support to joint judgement, drawing on design rationale approaches~\cite{maclean1991qoc}. Evidence requested to assess cost or feasibility should be linked to its source and available for inspection. Pairs should be able to invoke or defer this support as needed. Evaluation should examine whether it helps both collaborators articulate judgements and reconsider directions without imposing premature agreement or interrupting exploration.

\subsection{Supporting Joint Expression of Shared Meanings}

Eight pairs encountered AI responses that did not express shared design meanings (C2). Their navigation included bringing references to AI, explaining relationships and operating rules, and stating intended changes (N4--N6). P5's diagram reached the AI, but the hub coverage rule established in discussion still required explanation. The interviews also exposed a participation issue: P3 described relaying additions through the other collaborator, and P11 described one person typing both people's ideas. Suggestions for joint annotation and AI clarification questions (S2) concerned who could contribute to expressing a shared concept.

The interface helps account for this combination of shared understanding and individual expression. Both people could edit the canvas and access the discussion transcript, but each composed requests in a separate chat panel. A canvas selection became a single image; discussion accompanied it only when explicitly attached. The pair could therefore establish a relationship through conversation and drawing that was not specified in the submitted request. P5 provides process evidence for this distinction: the pair agreed on coverage before the diagram was submitted without the transcript or a written explanation of the rule. Shared material and the meaning needed to interpret it followed different paths into the request. Grounding between collaborators~\cite{clark1991grounding,tang1991observational} thus did not by itself ensure that the request expressed their understanding. This explains a need for expression support in our setting, while leaving open how the model would respond to other formulations.

Existing systems offer complementary mechanisms. Artinter lets clients and artists define concepts through visual examples, supporting a mutually understood vocabulary~\cite{chung2023artinter}. CoPrompt enables collaborators to share and refer to prompts and request assistance within a programming workspace~\cite{feng2024coprompt}. PromptHive provides a shared prompt library and facilities for cloning variants and comparing their outputs in educational content authoring~\cite{reza2025prompthive}. These approaches make references and prompt formulations available for collaborative use. Our findings motivate extending that support to the relationship between a shared visual concept and the request through which either collaborator asks AI to work on it.

Future systems could offer a shared draft for selected AI requests, where either person can annotate a reference region, explain a relationship, or add an intended change. The draft should retain links between these contributions and the relevant visual material. In a case like P5, both collaborators could inspect whether the request includes the coverage relationship as well as the diagram. AI-generated clarification questions could draw attention to unspecified relationships, with proposed interpretations available for correction before submission. Participation should remain flexible: P10 preferred viewing a partner's chat without editing it, so pairs should be able to invite a contribution to one request while retaining separate exploration. Evaluation should examine whether this reduces verbal relay and makes both people's intended contributions visible in the submitted request, alongside the correspondence of subsequent outputs to those intentions.

\subsection{Supporting Communication of Current Design Decisions}

The two C3 encounters concerned agreed changes that were absent from later AI responses. P11's output omitted the combined product developed through the other AI, while P8's output restored solar panels that the pair had removed. Pairs responded by supplying discussion or explicitly communicating the decision (N4, N6). The interviews broaden this observation into a support need: seven pairs suggested selective access to context across AIs (S3), including shared history, summaries, and sources. P8 wanted personal chats alongside a team chat, while P5 wanted to share only part of a history.

These encounters need to be interpreted in relation to how \duo{} handled context. Each chatbot retained its own chronological history, and requests received snapshots of selected canvas material and optionally the accumulated live transcript. Later canvas edits did not update previous attachments, and the two histories were not synchronised. Consolidation asked pairs to bring prior developments into a final account, making the relationship between earlier explorations and current decisions particularly relevant. P11's later request lacked the combined concept developed elsewhere; P8 supplied a current sketch, but the preceding AI history still contained the rejected solar-based design and the request did not explicitly state its removal. These records identify different input conditions under which a current decision was absent from an output. They do not establish which model processes produced the mismatch. The design opportunity is to help pairs inspect and communicate the status of relevant decisions when preparing a request.

Prior systems suggest mechanisms for connecting decisions with their context. Callisto links team discussions to notebook elements and their edit histories, making the reasoning around an artefact accessible from the artefact itself~\cite{wang2020callisto}. Memolet lets users select and manipulate reusable conversational memories when composing new AI interactions~\cite{yen2024memolet}. PaperPing, implementing Social-RAG, retrieves relevant signals from group interactions to contextualise AI-generated paper recommendations~\cite{wang2025socialrag}. SemanticCommit further supports inspecting potential conflicts when new intent is integrated into an existing specification, with proposed changes subject to user review~\cite{vaithilingam2025semanticcommit}. These systems suggest combining links to design history, context selection, and reviewable updates. Our encounters motivate making the status of a design decision explicit within such mechanisms, so that earlier possibilities can be distinguished from current commitments.

Future systems could link a design decision to its sketch region, supporting discussion, and relevant AI exchanges, while allowing collaborators to mark it as proposed, agreed, rejected, or reopened. A request preview could show which version and decisions will accompany the prompt. For P8, the selected sketch could include the decision to replace solar charging and its weight-related rationale; for P11, it could include the combined product representation developed through the other AI. Retrieval or summarisation could suggest relevant context and flag earlier statements that may conflict with an update, leaving the pair to review and correct the proposed changes. Contributors and unresolved differences should remain visible, and pairs should control which material enters personal or shared requests. Evaluation should examine whether this support helps later outputs reflect current decisions, how much additional record-keeping it requires, and whether collaborators can revise a decision without losing the reasons behind earlier alternatives.

\section{Limitations}

The study involved 12 pairs in short, co-located sessions with \duo{}. Twenty participants were aged 18--24, and all had design-course experience. Nine pairs were previously acquainted and three were formed for the study; familiarity was not compared. Longer projects and established professional teams may involve different ways of developing and communicating shared intentions. The structured synthesis, adaptation, and consolidation activities also shape the phase distributions reported here.

The categories and navigation matrix are interpretations of these sessions. N codes were linked to identified challenges, so the C--N distribution reflects this selection. Counts indicate coverage across pairs and phases, not request-level failure rates or strategy effectiveness. C3 rests on two encounters. Focusing on challenges may underrepresent routine or preventive work; the sixteen records without C codes warrant further analysis. The matrix's organisation by direction and design content offers a basis for examining further cases.

\section{Conclusion}

Our study of 12 pairs using \duo{} examined challenges and navigation in AI-supported collaborative design ideation. Pairs questioned the fit between AI proposals and shared purposes, clarified shared meanings missing from responses, and communicated agreed changes absent from later outputs. Their bridging work connected AI and shared design in two directions: jointly reworking, developing uses for, and assessing AI proposals, and bringing shared materials, meanings, and decisions to AI. These practices show how pairs shape their intentions through AI proposals and express those intentions as their design develops. They inform support for joint judgement, joint expression, and communication of evolving design context.

\FloatBarrier
\bibliographystyle{ACM-Reference-Format}
\bibliography{main}
\clearpage
\appendix
\onecolumn
\section{A Four-Phase Design Example}
\label{sec:appendix-phase-example}
\begin{figure}[H]
\centering
\includegraphics[width=\textwidth]{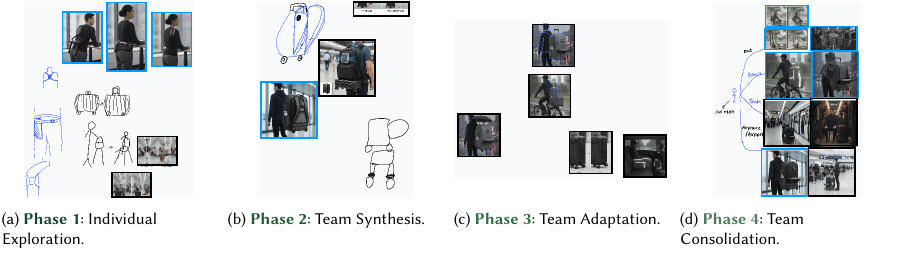}
\caption{P7's canvas across the four study phases: (a) Individual Exploration, (b) Team Synthesis, (c) Team Adaptation, and (d) Team Consolidation. The panels retain the sketches, generated images, blue outlines, and handwritten annotations visible in the source figure.}
\Description{Four side-by-side canvas snapshots show P7's work across the four phases, including black and blue sketches, backpack and wheeled-bag images, and blue outlines and annotations. The final panel arranges images alongside annotations concerning commuting contexts.}
\label{fig:p7-phases}
\end{figure}

Figure~\ref{fig:p7-phases} shows P7's shared canvas across the four study activities. The snapshots provide a visual reference for the procedure described in Section~\ref{sec:method}, showing sketches and generated images as the pair developed and consolidated their concept.

\clearpage
\section{Data Corpus}
\label{sec:appendix-data-sources}
Table~\ref{tab:data-sources} details the materials collected from the 12 paired sessions and the immediate post-session interviews.

\begin{table}[H]
\centering\small\setlength{\tabcolsep}{5pt}
\caption{Data corpus from 12 pairs. Study records document what participants and their AIs did; immediate post-session interviews capture participants' explanations and suggestions. Reconstructed transcripts cover both the design sessions and post-session interviews.}
\label{tab:data-sources}
\begin{tabular}{@{}Q{0.24\textwidth} Q{0.43\textwidth} Q{0.22\textwidth}@{}}
\toprule Data source & Material available for analysis & Coverage \\ \midrule
\rowcolor{Phase2!8}\textbf{Session screen recordings} & On-screen actions, verbal discussion, and the sequence of AI requests, generation, and canvas work. & 12 paired sessions; four activities per session. \\ \addlinespace[3pt]
\rowcolor{Phase2!8}\textbf{Reconstructed transcripts} & Transcriptions of session discussions and post-session interviews, manually checked for wording and speaker attribution. & Four activities per session and 12 joint post-session interviews. \\ \addlinespace[3pt]
\rowcolor{Phase2!8}\textbf{AI interaction records} & Recorded prompts, attached context, and AI-generated text and images within each collaborator's AI history. & Separate AI interactions throughout the session. \\ \addlinespace[3pt]
\rowcolor{Phase2!8}\textbf{Shared canvas materials} & Sketches, annotations, generated images, and their arrangement, interpreted alongside the recordings. & Shared design work across the four activities. \\ \addlinespace[3pt]
\rowcolor{N6!7}\textbf{Post-session interview transcripts} & Accounts of specific encounters, explanations of choices, and suggestions for improving collaboration with AI. & 12 joint interviews; approximately 15 minutes each. \\
\bottomrule\end{tabular}
\end{table}

\clearpage
\section{Detailed Qualitative Codebooks}
\label{sec:appendix-codebooks}
The tables below define the challenge (C), navigation (N), and perceived support (S) codes. C and N concern observed encounters and responses during the design activities; S concerns suggestions in pair interviews. C1--C3 address forming shared purposes, expressing shared meanings, and carrying agreed changes forward. N1--N3 concern developing AI proposals within shared design; N4--N6 concern communicating shared design to AI.

\begin{table}[H]
\centering\small\setlength{\tabcolsep}{5pt}
\caption{Codebook of challenges pairs encountered in connecting shared design with AI.}
\label{tab:challenges}
\begin{tabular}{@{}l Q{0.39\textwidth} Q{0.50\textwidth}@{}}
\toprule Code & Challenge & Observed difficulty \\ \midrule
\rowcolor{C1!7}\ccode{C1} & Unresolved Fit between AI Proposals and Shared Design Purposes & The pair raises a concrete question or trade-off about an AI proposal's purpose, use, or fit with shared constraints. Routine acceptance or elaboration alone is insufficient.\\ \addlinespace[2pt]
\rowcolor{C2!7}\ccode{C2} & Shared Design Meanings Not Expressed in AI Responses & An AI response fails to express an identifiable shared intention, as shown by a specific mismatch and clarification through words, sketches, or references. The intended meaning precedes the clarification. \\ \addlinespace[2pt]
\rowcolor{C3!7}\ccode{C3} & Agreed Design Changes Missing from Later AI Responses & A later AI output fails to reflect an already agreed update to the shared concept. Evidence connects the earlier state, the pair's change, and the subsequent mismatch. \\ \addlinespace[2pt]
\bottomrule\end{tabular}
\end{table}

\begin{table}[H]
\centering\small\setlength{\tabcolsep}{5pt}
\caption{Six navigation strategies. Each code identifies an observable response to a challenge. Definitions specify the action; examples illustrate it in the study. Several strategies may occur in one response sequence when supported by distinct actions.}
\label{tab:navigation-codebook}
\begin{tabular}{@{}l Q{0.26\textwidth} Q{0.63\textwidth}@{}}
\toprule Code & Navigation & Definition and study example \\ \midrule
\rowcolor{N1!7}\nshort{1} & Reworking AI Proposals for Shared Design & \textbf{Develop a revision or alternative to an AI proposal.} Collaborators alter, replace, or reorganise a proposed concept or feature through discussion, sketching, or further exploration. \emph{Example:} P1 questions the proposed finger-protection feature and develops a folding clip instead. The coded action is revising the concept, which can occur without another AI request. \\ \addlinespace[3pt]
\rowcolor{N2!7}\nshort{2} & Developing Uses for AI Proposals & \textbf{Develop how a proposed feature would be used.} Collaborators work out what it would do, when or how it would be used, or how it would contribute to their shared concept. \emph{Example:} P11 develops the proposed screen into a countdown display activated near a train platform. The coded action is developing its role or use. \\ \addlinespace[3pt]
\rowcolor{N3!7}\nshort{3} & Assessing AI Proposals against Shared Requirements & \textbf{Assess a proposal against a shared purpose or constraint.} Collaborators consider its value, limitations, or trade-offs in relation to requirements such as portability, cost, or power. \emph{Example:} P8 weighs suggested bottle functions against weight and solar-power limitations. The coded action is evaluating fit; a concrete revision is not required. A question or difficulty visualising a mechanism alone is insufficient without an articulated judgement against a purpose or constraint. \\ \addlinespace[3pt]
\rowcolor{N4!7}\nshort{4} & Bringing Shared Design Materials to AI & \textbf{Bring shared design material into an AI request.} A collaborator brings in shared discussion, a current sketch, or an output from the other collaborator's AI to provide context or a reference. \emph{Example:} P11b attaches the pair's discussion; P3b submits an image generated by P3a's AI. The coded action is transferring material. \\ \addlinespace[3pt]
\rowcolor{N5!7}\nshort{5} & Explaining Shared Design Meanings to AI & \textbf{Clarify what the AI should depict or explain.} A collaborator specifies the intended object, the meaning of a symbol, a spatial relationship, or an aspect of operation that the output needs to convey, using words or sketches. \emph{Example:} P5 explains what the diagram's letters represent and how alternating hubs serve adjacent stops. The coded action is explaining meaning. \\ \addlinespace[3pt]
\rowcolor{N6!7}\nshort{6} & Communicating Shared Design Decisions to AI & \textbf{Tell the AI what should change or stay.} A collaborator communicates an intended addition, removal, replacement, or retention in the next output. \emph{Example:} After the pair has rejected solar charging, P8a asks the AI to remove the solar panels. The coded action is communicating the revision. \\
\bottomrule\end{tabular}
\end{table}

\paragraph{Coding boundary: C1 and N3.} P8b's remark that an AI-proposed storage compartment \dquote{would defeat the purpose of it being collapsible} supports both codes: it identifies unresolved fit (C1) and assesses the addition against a shared portability requirement (N3). The same utterance supports both analytical roles. P8a's subsequent suggestion to develop one collapsible variant and one non-collapsible variant supports N1 through a distinct design action. In contrast, P2b's \dquote{I can't see how that'll work}, referring to a proposed silicone strap, identifies a C1 difficulty but does not alone supply a criterion-based assessment. The pair's development of a cap-and-slot alternative supports N1; this record was not assigned N3. N3 requires an articulated assessment against a shared purpose or constraint.

\paragraph{Linking navigation to challenges.} After developing a proximity condition for the screen (N2), P11a sent that condition to AI. The request carried a newly formed use into the next interaction. As it addressed no previously identified expression mismatch, it received no additional N4--N6 assignment in the challenge-linked mapping. In P3, transferring an image (N4) supplied a reference, followed by clarification of rack placement (N5). These examples show how the assignments relate observable actions to the challenge being addressed; Table~\ref{tab:navigation-counts} counts those links.

\begin{table}[H]
\centering\small\setlength{\tabcolsep}{5pt}\renewcommand{\arraystretch}{1.15}
\caption{Detailed definitions and interview examples of the three perceived support codes. Codes may overlap when suggestions contain distinct support needs.}
\label{tab:support-codebook}
\begin{tabularx}{\textwidth}{@{}l Q{0.23\textwidth} Y@{}}
\toprule Code & Support need & Definition and interview example \\ \midrule
\rowcolor{S1!18}\scode{1} & Jointly assessing and revisiting AI proposals & Provide grounds for assessing design directions or ways to return to and revise them. \emph{Example:} A P9 participant requested studies and numbers to assess a concern; P10b proposed revisiting an earlier point in a shared timeline to revise the concept and its AI context. \\ \addlinespace[3pt]
\rowcolor{S2!13}\scode{2} & Jointly expressing shared design intentions to AI & Enable collaborators to contribute more directly to specifying or clarifying what an AI should convey. \emph{Example:} P4a proposed that both people could circle regions of the same image to request changes; P6b wanted the AI to ask clarifying questions. \\ \addlinespace[3pt]
\rowcolor{S3!13}\scode{3} & Selective access to context across separate AIs & Make selected AI history and shared discussion available across personal and shared work, including control over what is shared. \emph{Example:} P5b wanted to share part of their AI history directly with the other collaborator; P8a wanted personal and team chats alongside one another. \\
\bottomrule
\end{tabularx}
\end{table}

\end{document}